\documentclass[aps,amsfonts,amsmath,prd,preprint,tightenlines,showpacs,nofootinbib]{revtex4}

\def\be{\begin{equation}}
\def\ee{\end{equation}}
\def\bea{\begin{eqnarray}}
\def\eea{\end{eqnarray}}
\def\bml{\begin{subequations}}
\def\blea{\bml\begin{eqnarray}}
\def\elea{\end{eqnarray}\end{subequations}}

\def\teq{t_{\text{eq}}}
\def\zeq{z_{\text{eq}}}
\def\Mmin{M_{\text{min}}}
\def\lmin{l_{\text{min}}}
\def\fcoll{f_{\text{coll}}}
\def\fstar{f_{\text{star}}}
\def\fesc{f_{\text{esc}}}

\begin{document}

\title{Reionization from cosmic string loops}

\author{Ken D. Olum}
\email{kdo@cosmos.phy.tufts.edu}
\author{Alexander Vilenkin}
\email{vilenkin@cosmos.phy.tufts.edu}
\affiliation{Institute of
Cosmology, Department of Physics and Astronomy, Tufts University,
Medford, MA  02155}

\begin{abstract}

Loops formed from a cosmic string network at early times would act as
seeds for early formation of halos, which would form galaxies and lead
to early reionization.  With reasonable guesses about astrophysical
and string parameters, the cosmic string scale $G\mu$ must be no more
than about $3\times 10^{-8}$ to avoid conflict with the reionization
redshift found by WMAP.  The bound is much stronger for superstring
models with a small string reconnection probability.  For values near
the bound, cosmic string loops may explain the discrepancy between the
WMAP value and theoretical expectations.

\end{abstract}


\maketitle

\section{Introduction}

%%alias=superstrings=hep-th/0204074,hep-th/0312007,hep-th/0312067%%

Cosmic strings are linear topological defects which may have been
formed in the early universe via phase transitions \cite{Kibble} or
through brane annihilation in superstring theory \cite{superstrings}.
Once formed, cosmic strings exist at any time in a ``network''
of loops and infinite strings.  The network evolves in a scaling
regime in which any linear measure of the network properties is a
constant fraction of the horizon size.  This dynamic is maintained by
the production of loops via reconnection on long strings, and
the subsequent evaporation of loops by gravitational radiation.

The energy scale of strings can be given by the dimensionless number
$G\mu$, where $\mu$ is the linear energy density on the string and $G$
is Newton's constant.  In the early days, cosmic strings were a
candidate for the source of structure in the universe, either through
accretion of matter onto string loops or onto the wakes of moving
strings.  This scenario, which required $G\mu\gtrsim 10^{-6}$, has
been conclusively ruled out by microwave background observations,
and current observations limit $G\mu$ to be less than about $2\times
10^{-7}$ \cite{astro-ph/0604141}.

Nevertheless, even at smaller $G\mu$, there will be some amount of
structures formed by cosmic strings, in particular through accretion
around loops. Localized seeds like loops can form nonlinear
structures at very early times. This could result in early star
formation and in reionization of the universe.  Even a small
percentage of baryons in stars leads to reionization. The time of
reionization is constrained by WMAP observations
\cite{astro-ph/0603449}, yielding a bound on the string parameter
$G\mu$.

%%alias=AT=Phys.Rev.Lett.54.1868,Phys.Rev.D40.973%%
%%alias=BB=Phys.Rev.Lett.60.257,Phys.Rev.Lett.63.2776%%

The idea that strings could cause reionization has been discussed
by a number of authors \cite{Rees,Hara,Avelino,Levon}. All these
papers assumed that the effect of strings on structure formation is
mostly through wakes formed behind rapidly moving long strings. The
effect of loops was neglected because the loops were assumed to be too
small and too short-lived. The resulting bound on $G\mu$ was
$G\mu\lesssim 10^{-6}$. Here, we reconsider these results in the light
of recent cosmic string simulations.

The formation of stars by string loops depends on the loop sizes being
large enough to accrete sufficient matter for a galaxy.  Early
simulations \cite{AT} found loops at a large fraction of the horizon
size, in accordance with theoretical expectations. However, later
simulations \cite{BB,Phys.Rev.Lett.64.119} found loops at much smaller
sizes, essentially the minimum resolution of the
simulations.\footnote{More recent simulations \cite{Bouchet,Shellard}
found evidence of loop scaling, but the loop sizes were still very
small, less than 0.001 of the horizon.} This recently led us, in
collaboration with Vitaly Vanchurin, to develop a simulation
\cite{gr-qc/0501040} which does not have a minimum resolution size.
We found \cite{gr-qc/0511159} that loops were formed with lengths of
about 1/10 of the simulation time (which plays the role of the horizon
in our flat-space simulation).  This pattern established only after a
long transient period dominated by very small loops, comparable to the
initial scale of the network. We believe that it was this transient
regime that was observed in earlier simulations. String evolution in
the expanding universe is expected to be qualitatively similar,
although with a somewhat different ratio of loop size to cosmic time.
Here, we will show that large loop sizes could indeed lead to early
reionization, yielding a stringent bound on $G\mu$.

\section{Loop distribution}

The string loops of interest to us here will be those which
formed during the radiation era but have not yet decayed at
$\teq$. The energy density of loops that were chopped off the network
in one Hubble time is comparable to the energy density of long
strings. However, the loop energy redshifts like matter, while the
long string energy redshifts like radiation. So, if loops are long and
live much longer than a Hubble time, they dominate the energy of the
network and play the dominant role in structure formation.

We will assume there is a scaling process
of production, which means that
\be
n(l,t) = t^{-5} f(x) \qquad \text{with $x=l/t$}
\ee
where $n(l, t) dl$ is the number density of loops produced
with length between $l$ and $l+dl$ in unit time in unit spatial
volume.  We will take the loop production to be given by a power law
distribution up to a certain fraction of the cosmic time,
\be\label{eqn:f}
f(x) = A x^{-\beta} \qquad \text{for $x<\alpha$}
\ee
and zero otherwise.  The scaling network is characterized by some
inter-string distance $d(t) =\gamma t$, defined so that the density in
long strings is $\rho_\infty =\mu/d^2$.  Conservation of energy 
then gives
\be\label{eqn:fnorm}
\int_0^\infty x f(x) dx= \frac{1}{\gamma^2}\left(1-\langle
v^2\rangle\right)
\ee
so from Eq.\ (\ref{eqn:f}),
\be
A = \frac{2-\beta}{\alpha^{2-\beta}\gamma^2}\left(1-\langle v^2\rangle\right)
\ee
Here, $\langle v^2\rangle$ is the square of string velocity averaged
along the length of long strings.

A loop of length $l$ evaporates by gravitational radiation in time
$l/(\Gamma G\mu)$, where $\Gamma$ is a number of order 50.  Thus if we
consider loops with $l \gg \Gamma G\mu t$ they will not have undergone
significant evaporation.
The length distribution of such loops in the radiation era is
then given by
\be
N(l, t) = \frac{g}{t^{3/2}l^{5/2}}\int_0^\infty x^{3/2} f(x) dx\,,
\ee
where $g = \sqrt{1-v_i^2}$ and $v_i$ is the initial center of
mass velocity of the loops.
Using Eqs.\ (\ref{eqn:f},\ref{eqn:fnorm}),
for $l < \alpha t$,
\be\label{eqn:N}
N(l, t)= \frac{{\cal N}}{t^{3/2}l^{5/2}}
\ee
with
\be
{\cal N} = \frac{g(2-\beta)\sqrt{\alpha}}{(5/2-\beta)\gamma^2}
\left(1-\langle v^2\rangle\right)
\ee

In \cite{gr-qc/0511159} we found loops emitted with significant substructure, so
that their center of mass velocities are low and $g\sim1$.  The
specific simulations of \cite{gr-qc/0511159} found $\alpha\approx
0.1$, $\gamma\approx 0.04$, $\beta\approx 1.6$, $\langle v^2\rangle =
0.4$, so
\be
{\cal N}_{flat}\sim50\,.
\ee
But since these simulations were done in flat space, there is no
reason to think that this value is correct for the
radiation-dominated universe.  A somewhat better motivated 
estimate can be
obtained by assuming that the parameters $\gamma$ and $\langle
v^2\rangle$ characterizing the long string network have been correctly
determined in the early simulations \cite{BB,Phys.Rev.Lett.64.119},
and that the loop sizes are comparable to the inter-string distance,
as in flat space, after the true scaling regime sets in.
Then $\alpha\sim\gamma\sim 0.25$, $\langle v^2\rangle \sim
0.4$, and
${\cal N}\sim 2$. {We shall assume that 
\be
{\cal N}\gtrsim 2
\label{Nrange}
\ee
in what follows.  A more accurate estimate must await long-duration
expanding-universe string simulations.}

\section{Formation of halos and reionization}

At the time of matter-radiation equality, $\teq$, loops start to
accrete dark matter.  In about one Hubble time, the mass of a
loop-seeded halo becomes comparable to that of the loop itself, so
that the subsequent decay of the loop has little effect on the
accretion process. At some $t > t_{eq}$, the halo seeded by a loop of
length $l$ will have
accreted mass
\be
M(l)\sim \mu l \left(\frac{t}{\teq}\right)^{2/3} = \mu l \frac{1+\zeq}{1+z}
\ee
in cold dark matter.  The number density of halos
formed around loops that existed at $\teq$ will be
\be
n(l, t) \sim \frac{{\cal N}}{\teq^{3/2}l^{5/2}}
\left(\frac{1+z}{1+\zeq}\right)^3
\ee
Once the halo exceeds the Jeans mass, it will start to accrete baryons
as well as dark matter.  After recombination, the Jeans mass
(including both dark matter and baryons) is about
$10^5 M_\odot$, but a halo must exceed some larger threshold
$\Mmin$ in order to be able to cool and form stars.  
Thus only loops with length at least
\be
\lmin = \frac{\Mmin (1+z)}{\mu(1+\zeq)}
\ee
will form luminous galaxies by redshift $z$.

The total mass density of such galaxies is
\be
\mu \frac{1+\zeq}{1+z}\int_{\lmin} n(l, t) l dl = 
2\frac{{\cal N}\mu}{\teq^{3/2}\lmin^{1/2}}
\left(\frac{1+z}{1+\zeq}\right)^2
= 2\frac{{\cal N}\mu^{3/2}}{\teq^{3/2}\Mmin^{1/2}}
\left(\frac{1+z}{1+\zeq}\right)^{3/2}.
\ee
The total mass density of the matter-dominated universe
is $1/(6\pi G t^2)$, so the fraction of
collapsed matter in halos larger than $\Mmin$ is
\be
\fcoll = 12\pi\frac{{\cal N} G\mu^{3/2}\teq^{1/2}}{\Mmin^{1/2}}
\left(\frac{1+\zeq}{1+z}\right)^{3/2}\,.
\ee
With $\zeq = 5000$ and $\teq = 10^{12}$s, we get
\be
\fcoll\approx 6\times 10^{15}
\frac{{\cal N}\left(G\mu\right)^{3/2}}{(1+z)^{3/2}}
\left(\frac{\Mmin}{M_\odot}\right)^{-1/2}
\ee

Now, for a significant amount of star formation we need a halo with a
virial temperature above $10^4$K to allow atomic hydrogen cooling.
This requires\footnote{This is derived by
setting $T_{vir}$ in Eq.(86) of \cite{astro-ph/0603360} to $10^4K$,
with mean molecular weight 1.2 (atomic gas).}
\be
\Mmin \sim 10^9(1+z)^{-3/2}M_\odot,
\ee
so
\be
\fcoll\approx 2\times 10^{11}
\frac{{\cal N}\left(G\mu\right)^{3/2}}{(1+z)^{3/4}}.
\ee

Of those baryons in halos, some fraction
\be\label{eqn:fstar}
\fstar\sim 0.1
\ee
will participate in star formation.  The
number of ionizing photons produced per baryon is about
\be\label{eqn:nion}
4\times 10^3 - 10^5
\ee
where the higher number 
corresponds to metal-free stars. (For an up-to-date review of the
physics of reionization, see \cite{astro-ph/0603360}.)
The metallicity is likely to
grow rather quickly as the first stars explode as supernovae, hence
we are going to use the more conservative estimate corresponding to
the lower number in (\ref{eqn:nion}).  Some fraction
\be\label{eqn:fesc}
\fesc\sim 0.1
\ee
of the ionizing photons escape from their galaxies.  The total ratio
of intergalactic ionizing photons to baryons is thus about
\be
4\times 10^3\fcoll\fstar\fesc
\ee
The characteristic recombination time for ionized hydrogen is
\be
\tau_{\text{rec}}\sim \frac{50}{(1+z)^{3/2}C}\,t
\ee
where $C= \langle n_H^2\rangle/\bar n_H^2\sim10$ is the ``clumpiness factor''.
(This follows From Eq.\ (120) of \cite{astro-ph/0603360}.)
Thus at redshifts $z\sim 15$ of interest here the universe will not be
completely reionized until we have emitted some number
\be\label{eqn:ni}
n_i\sim10
\ee
of photons per baryon.  Thus reionization takes place when
\be
4\times 10^3\fcoll\fstar\fesc= n_i\,.
\ee
Complete reionization is ruled out by the third-year WMAP data
for $z>13.6$ (one sigma) \cite{astro-ph/0603449}.  Thus we must have
$\fcoll \lesssim 3 \times 10^{-4} n_i \fstar^{-1}\fesc^{-1}$ at this
redshift, which means that
\be\label{eqn:bound}
G\mu\lesssim 4\times 10^{-10} ({\cal N}\fstar\fesc/n_i)^{-2/3}.
\ee
With the estimates of Eqs.\ (\ref{Nrange},\ref{eqn:fstar},\ref{eqn:fesc},\ref{eqn:ni}), the bound is $G\mu\lesssim 3\times 10^{-8}$.

\section{Discussion}

A cosmic string network can produce loops that act as seeds for the
formation of some small galaxies at early times.  These galaxies
will lead to reionization at larger redshifts than allowed by WMAP
data, unless the string energy scale obeys the bound of Eq.\
(\ref{eqn:bound}). This bound relies mostly on the general
argument, confirmed by simulations \cite{gr-qc/0511159}, that strings
are formed at a substantial fraction of the horizon size, rather than
tiny scales set by gravitational back reaction. 

The bound (\ref{eqn:bound}) is to be compared with constraints on
$G\mu$ coming from other phenomena. As we already mentioned, the
current bound from CMB observations is $2\times 10^{-7}$
\cite{astro-ph/0604141}. The bounds from millisecond pulsar timing
\cite{Damour} and from nucleosynthesis considerations
\cite{gr-qc/0511159} are both $G\mu\lesssim 10^{-7}$. If the
parameter ${\cal N}$ in Eq.\ (\ref{eqn:bound}) is in the assumed range
(\ref{Nrange}), and given the assumptions of Eqs.\ (\ref{eqn:fstar},
\ref{eqn:fesc}), the reionization bound is 
\be
G\mu \lesssim 3\times 10^{-8},
\label{bound}
\ee
somewhat stronger than presently available bounds.
We emphasize, however, that precise values of ${\cal N}$,
$\fstar$, $\fesc$, and $n_i$ are presently unknown and the bound
(\ref{bound}) should be regarded as preliminary.

If we consider a small intercommutation probability $p$, as appears in
cosmic superstring models \cite{Jackson}, the density of strings will
be increased for a given $G\mu$, and so the bound will become more
stringent.  A reasonable conjecture is that $p$ does not affect the
loop sizes, but the overall density is increased by a factor $1/p$
\cite{Damour,Sakellariadou}.  Then ${\cal N}\propto
1/p$, and the limit on $G\mu$ is proportional to $p^{2/3}$.

We note finally that a value of $G\mu$ near the reionization bound
(\ref{bound}) may explain the apparent discrepancy
\cite{Alvarez,Madau} between the three-year WMAP data suggesting
reionization at $z\sim 11$ and the star formation theory indicating
that the formation of a sufficient number of stars at such early
redshift is unlikely in the standard scenario. Strings with $G\mu\sim
3\times 10^{-8}$ may account for early star formation, although such
strings will play little role in structure formation at later epochs.

\section{Acknowledgments}

We would like to thank Avi Loeb for very helpful comments.  This work
was supported in part by the National Science Foundation under grants
0353314 and 0457456.

\end{document}